\documentclass[11pt]{article}
\usepackage[preprint]{acl}
\usepackage{times}
\usepackage{latexsym}
\usepackage[T1]{fontenc}
\usepackage[utf8]{inputenc}
\usepackage{microtype}
\usepackage{inconsolata}
\usepackage{graphicx}
\usepackage{amsmath}
\usepackage{amssymb}
\usepackage{bbm}
\usepackage{subcaption}
\usepackage{float}
\usepackage{booktabs}
\usepackage{tabularx}
\usepackage{array}
\usepackage{xcolor}
\usepackage{multirow}
\usepackage{colortbl}
\usepackage{algorithm}
\usepackage{algpseudocode}
\usepackage{enumitem}
\usepackage{tikz}
\usepackage{hyperref}
\usetikzlibrary{shapes,arrows,positioning}
\usepackage{pifont}
\usepackage{tcolorbox}

\newcommand{\silobench}{\textsc{Silo-Bench}}

\definecolor{lightblue}{RGB}{230,240,255}
\definecolor{lightgreen}{RGB}{230,255,230}
\definecolor{lightyellow}{RGB}{255,255,230}
\definecolor{red}{RGB}{220,40,30}
\definecolor{BrightOrange}{RGB}{255,140,0}
\definecolor{Gold}{RGB}{255,200,0}
\definecolor{Emerald}{RGB}{0,160,80}
\definecolor{Cobalt}{RGB}{0,120,215}
\definecolor{Sapphire}{RGB}{90,60,170}
\definecolor{Amethyst}{RGB}{160,80,200}
\definecolor{framegray}{RGB}{170,170,170}

\usepackage{listings}
\title{%
  \raisebox{-0.3em}{\includegraphics[height=1.7em]{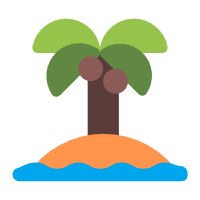}}%
  ~\silobench{}: A Scalable Environment for Evaluating Distributed\\Coordination in Multi-Agent LLM Systems%
}

\author{
  \textbf{Yuzhe Zhang}$^{1}$ \quad
  \textbf{Feiran Liu}$^{1}$ \quad
  \textbf{Yi Shan}$^{1}$ \quad
  \textbf{Xinyi Huang}$^{1}$ \quad
  \textbf{Xin Yang}$^{2}$ \quad
  \textbf{Yueqi Zhu}$^{1}$ \\
  \textbf{Xuxin Cheng}$^{4}$ \quad
  \textbf{Cao Liu}$^{4}$ \quad
  \textbf{Ke Zeng}$^{4}$ \quad
  \textbf{Terry Jingchen Zhang}$^{3,5\dagger}$ \quad
  \textbf{Wenyuan Jiang}$^{3\dagger}$\thanks{$^{\dagger}$Corresponding Author. Email: wenyjiang@ethz.ch} \\[0.6em]
  $^{1}$\,Beijing University of Technology, Beijing, China \\
  $^{2}$\,Zhejiang University, Hangzhou, China \quad $^{3}$\,ETH Z\"{u}rich, Switzerland \\
  $^{4}$\,Meituan LongCat Interaction Team \quad $^{5}$\,Vector Institute for Artificial Intelligence
}
\begin{document}
\maketitle
\begin{abstract}
Large language models are increasingly deployed in multi-agent systems to overcome context limitations by distributing information across agents. 
Yet whether agents can reliably \textbf{compute} with distributed information, rather than merely exchange it, remains an open question. 
We introduce \silobench{}, a role-agnostic benchmark of 30 algorithmic tasks across three communication complexity levels, evaluating 54 configurations over 1,620 experiments. 
Our experiments expose a fundamental \textbf{Communication-Reasoning Gap}: agents spontaneously form task-appropriate coordination topologies and exchange information actively, yet systematically fail to synthesize distributed state into correct answers. 
The failure is localized to the reasoning-integration stage where agents often acquire sufficient information but cannot integrate it. 
This coordination overhead compounds with scale, eventually eliminating parallelization gains entirely. 
These findings demonstrate that naively scaling agent count cannot circumvent context limitations, and \silobench{} provides a foundation for tracking progress toward genuinely collaborative multi-agent systems. 
The code is available at \href{https://github.com/jwyjohn/acl26-silo-bench}{https://github.com/jwyjohn/acl26-silo-bench}.
\end{abstract}

\section{Introduction}
\label{sec:introduction}

The rapid advancement of Large Language Models (LLMs) has demonstrated remarkable capabilities in individual inference and generation tasks \citep{ai2023gpt, touvron2023llama, liu2024deepseek}. 
However, as the scale and complexity of real-world problems continue to grow, a fundamental bottleneck has emerged: the limited context window of a single model restricts its ability to process global information \cite{li2024loogle,chen2024long,an2024eval}. 
Even with recent progress in extending context lengths to millions of tokens \citep{reid2024gemini}, the quadratic cost of attention \citep{ratner2023parallel, yen2024long} makes centralized processing increasingly impractical for truly large-scale tasks.

\begin{figure}[t]
\vspace{1.0em} 
\centering
\includegraphics[width=1.0\columnwidth]{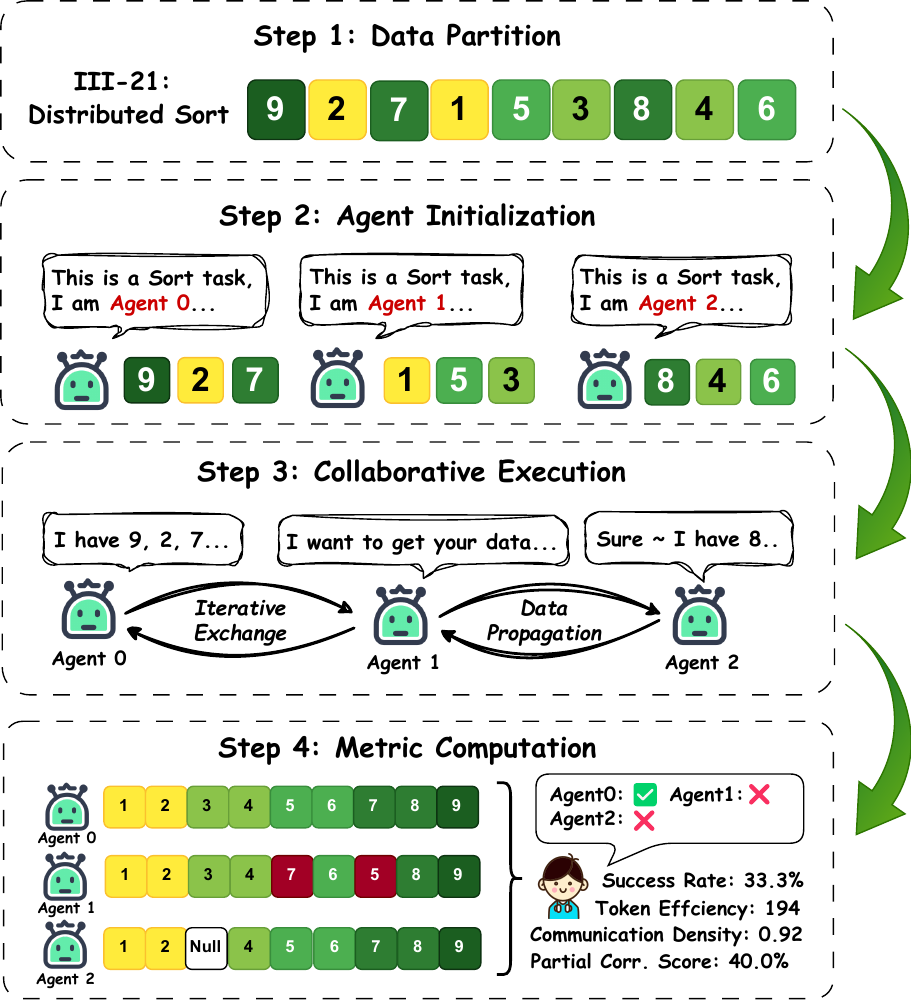}
\caption{Pipeline of \silobench{}. Global information is partitioned across \(N\) agents, each holding only local data. Agents must communicate through the provided protocol to reconstruct global truth. Success requires effective collaboration strategies. This is an example of the III-21 Distributed Sort (Appendix \ref{app:tasks}.)}
\label{fig:overview}
\end{figure}

Multi-Agent Systems (MAS) offer a compelling architectural paradigm to address this scalability challenge \citep{zhang2024exploring, wang2024rethinking}. 
By distributing global information across multiple agents that collaborate to compute results, MAS can theoretically overcome the token limitations of single models \cite{liuagentbench}. 
This distributed approach mirrors successful patterns in traditional computing, from MapReduce to modern distributed databases, where data partitioning and coordinated computation across nodes achieve scales unattainable by a single machine \cite{dean2008mapreduce}. 
In the realm of large models, we define the scenario where an individual agent has access only to partial information, thereby necessitating coordination to resolve token constraints, as information silos. 
However, a critical question remains underexplored: Can current LLM-based agents effectively collaborate within information silos to compute a globally correct answer \cite{qian2024chatdev,qian2023communicative,liuagentbench}? 
Existing multi-agent benchmarks either prescribe fixed communication structures \citep{li2023camel, wu2024autogen, hong2023metagpt} or focus on social simulation rather than computational collaboration \citep{park2023generative, lan2024llm}. 
These approaches often introduce inductive bias into the agents' final outputs \citep{baltaji2024conformity}. 
For instance, if an agent is assigned the role of a "doctor", it may exhibit poor performance in artistic domains, which contradicts the goal of developing general-purpose agents \cite{an2024eval,qian2024chatdev,qian2023communicative,liuagentbench}. 
Furthermore, most benchmarks to date target specific tasks \citep{deng2024mobile, chen2024llmarena, gioacchini2024agentquest} and fail to address a significant gap in our understanding: whether logic-based models can autonomously discover and execute effective coordination strategies for distributed computing problems. 
Addressing this gap is a key objective for future AGI evaluation. 
To bridge this gap, we propose \silobench{}—a pioneering benchmark for evaluating free-form communication and collaboration in multi-agent LLM systems \cite{liuagentbench}. 
In summary, our contributions are as follows:
\begin{itemize}[leftmargin=*]
\item \textbf{We introduce \silobench{}, a role-agnostic configurable environment for evaluating distributed coordination under information silos.} 
Unlike static test suites that prescribe fixed roles and communication scripts, our framework can generate unlimited evaluation instances while providing high-level task hints and allowing observation of whether agents can translate structural understanding into effective coordination protocols \cite{presstrain,liuagentbench}. 
\item \textbf{We conduct the largest systematic study of multi-agent collaboration to date by instantiating 54 representative configurations.} 
Spanning diverse protocols and computing paradigms \citep{zhao2024electoral, islam2024mapcoder}, we propose a multi-dimensional metric suite to comprehensively quantify the trade-off between task success rate, token consumption, and communication density.
\item \textbf{We expose critical scalability limitations and the \emph{Communication-Reasoning Gap} in current LLMs.} 
Our results reveal that while agents can spontaneously discover task-appropriate communication topologies, they fail to translate effective coordination into correct distributed computation. 
This disconnection, coupled with inefficient information synthesis, causes performance to collapse as task complexity increases and the agent scale expands.
\end{itemize}

\section{Related Work}
\label{sec:related_work}

\begin{figure*}[h]
    \centering
    \includegraphics[width=1.0\textwidth]{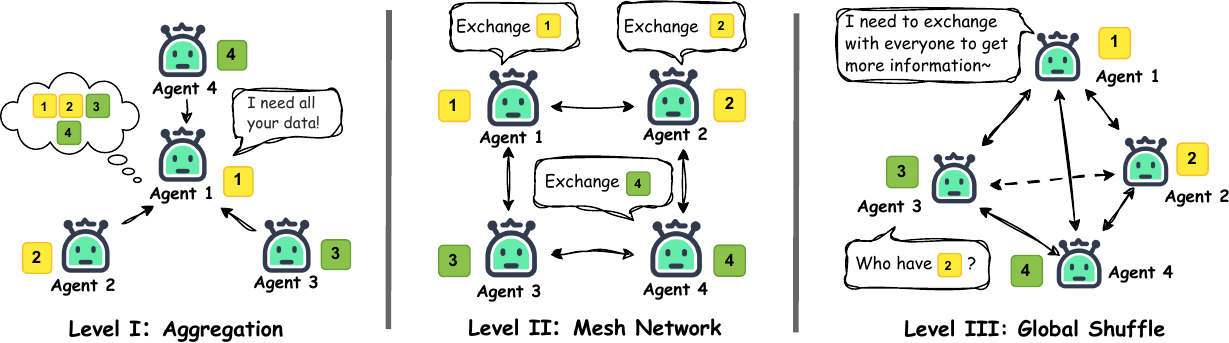}
    \caption{Three complexity levels in \silobench{} characterized by their communication patterns. \textbf{Level I (Aggregation)}: A central agent collects data from all peers via a star topology. \textbf{Level II (Mesh Network)}: Agents exchange information with immediate neighbors through pairwise communication. \textbf{Level III (Global Shuffle)}: All agents must communicate with every other agent, requiring full mesh connectivity.}
    \label{fig:task_levels}
\end{figure*}

\paragraph{Context Limitations and Distributed Reasoning.}
The finite context window of LLMs constitutes a fundamental bottleneck for processing large-scale information. While recent advances have extended context lengths to millions of tokens \citep{reid2024gemini,liuworld}, the quadratic computational complexity of attention mechanisms makes centralized processing increasingly resource-intensive and prone to ``lost-in-the-middle'' phenomena \citep{liu2024lost}. 
Although Retrieval-Augmented Generation (RAG) offers a palliative solution \citep{wang2024m,islam2024open}, it often fractures global context, struggling with tasks that require holistic reasoning across disjoint segments. 
Existing benchmarks like SCROLLS \citep{shaham2022scrolls}, LongBench \citep{bai2024longbench}, and \(\infty\)Bench \citep{zhang2024infty} effectively evaluate single-agent retrieval but overlook the paradigm of \textit{distributed collaboration}. 
We posit that overcoming the context barrier requires shifting from centralized attention to collaborative computation, where agents act as distributed processors, to digest partitioned information and synthesize global insights, which is a capability currently unmeasured by standard long-context evaluations.

\paragraph{Multi-Agent Architectures and Role-Agnosticism.}
The paradigm of orchestrating multiple LLM agents has evolved from simple role-playing to complex problem-solving frameworks. 
Foundational works like CAMEL \citep{li2023camel} and MetaGPT \citep{hong2023metagpt} utilize \textit{role-specialized agents} (e.g., assigning ``Manager'' or ``Coder'' personas) embedded within fixed hierarchical or waterfall workflows. 
While effective for domain-specific tasks like software engineering \citep{islam2024mapcoder} and interactive image generation \citep{ma2025talk2image}, these approaches entangle the agents' reasoning capabilities with semantic role priors, making it difficult to isolate the contribution of the communication architecture itself. 
Other efforts, such as debate-based systems \citep{du2023improving} or Mixture-of-Agents \citep{wangmixture,su2026daao}, often prescribe static topological constraints that limit dynamic information flow. 
\textbf{We introduce \silobench{}, a role-agnostic configurable environment with task-structural guidance for evaluating distributed coordination under information silos.} 
Unlike static test suites that prescribe fixed roles and communication scripts, our framework dynamically generates unlimited evaluation instances while providing high-level task hints, allowing observation of whether agents can translate structural understanding into effective coordination protocols \cite{presstrain,liuagentbench}.

\section{\silobench{}}
\label{sec:benchmark}

This section presents the architecture of \silobench{}, a configurable environment for evaluating multi-agent collaboration under information silos. 
Each configuration is defined by three orthogonal dimensions: agent scale $N$, communication protocol $P$, and language model $\mathcal{M}$. We describe the task space, evaluation metrics, and execution pipeline.

\subsection{Task Space}
\label{sec:task_space}

A central design goal of \silobench{} is to ground task difficulty in principled communication complexity theory, so that observed performance gaps can be attributed to coordination demands rather than ad hoc task choice. 
The theoretical foundation for analyzing distributed computation costs dates back to Yao's seminal work on communication complexity \citep{yao1979some}, which established the framework for quantifying the minimum bits required for distributed parties to compute a function. 
Building on this foundation, we categorize tasks by their optimal communication complexity:

\begin{equation}
\tau_k = (f_k, \mathcal{X}_k, y^*_k)
\label{eq:task_def}
\end{equation}
where $f_k$ specifies the computational function, $\mathcal{X}_k$ is the global input data, and $y^*_k$ is the ground-truth answer. 
Tasks are organized into three levels based on their optimal communication complexity (complete task specifications are provided in Appendix~\ref{app:tasks}).

\paragraph{Level I: Aggregation ($\mathcal{O}(N)$ communication).}
As illustrated in Figure~\ref{fig:task_levels} (left), these tasks exhibit embarrassingly parallel structure followed by reduction. 
Each agent processes its local shard independently, producing intermediate results aggregated through associative operations (e.g., \texttt{max}, \texttt{sum}, \texttt{xor}). 
The optimal topology is a star or tree structure where one agent collects all partial results. 
Representative tasks include global maximum (LC-414: ``Third Maximum Number''), distributed voting (LC-169: ``Majority Element''), and word frequency counting (LC-2085).

\paragraph{Level II: Mesh Network ($\mathcal{O}(N)$ communication).}
As shown in Figure~\ref{fig:task_levels} (center), these tasks exhibit spatial locality: agent $i$'s computation depends primarily on neighboring agents $i-1$ and $i+1$. 
Information propagates through a structured mesh via pairwise exchanges, with optimal topology being a linear chain requiring $N-1$ point-to-point exchanges. 
Representative tasks include prefix sum (LC-1480), moving average (LC-346: ``Moving Average from Data Stream''), and trapping rain water (LC-42).

\paragraph{Level III: Global Shuffle ($\mathcal{O}(N \log N)$ to $\mathcal{O}(N^2)$ communication).}
As depicted in Figure~\ref{fig:task_levels} (right), these tasks feature irregular, potentially all-to-all communication patterns where any agent's output may depend on information from any other agent. 
The range $\mathcal{O}(N \log N)$--$\mathcal{O}(N^2)$ spans from the classical lower bound for distributed reorganization to the full-consensus cost imposed by our evaluation criterion, where every agent must output the complete global answer. 
Representative tasks include distributed sorting (LC-912), graph connectivity (LC-323), and matrix multiplication (LC-311).

\subsection{Task Construction Pipeline.}
LeetCode problems serve solely as algorithmic inspiration and we \textit{do not} transform raw LeetCode data. 
For each task \emph{category} (e.g., ``Global Maximum''), we independently implement a Python generator that programmatically produces random inputs and exact ground-truth answers. 
A \emph{task instance} is one concrete input--output pair drawn from this generator under a fixed $(N, P, \mathcal{M})$ configuration, where $N$ is the agent scale, $P$ the communication protocol, and $\mathcal{M}$ the language model, and a fixed random seed, ensuring reproducibility while allowing unlimited fresh instances.

To illustrate: for Level-I Global Maximum (inspired by LC-414, the ``Third Maximum Number'' problem), given agent count $N$ and per-agent shard size $k$, the generator (i) samples $N \times k$ integers uniformly at random, (ii) partitions them into $N$ equal shards $\mathcal{X}_1,\ldots,\mathcal{X}_N$, (iii) computes $y^* = \max(\mathcal{X})$, and (iv) records the fixed seed for reproducibility. 
Each agent receives only its local shard $\mathcal{X}_i$ and must coordinate to determine $y^*$. 
This pipeline generalizes directly to all 30 tasks: the generator encodes the task-specific function $f_k$, scales global input size proportionally with $N$ to maintain constant per-agent workload, and produces exact ground-truth answers enabling fully objective evaluation.

\subsection{Evaluation Metrics}
\label{sec:metrics}

We define four complementary metrics to capture both \textit{what} agents achieve and \textit{how} they coordinate. 
Let $\hat{y}_i$ denote agent $i$'s submitted answer, and let $m_i$ denote the total number of messages successfully transmitted outward by agent $i$ during the entire collaboration.

\paragraph{Success Rate ($\mathcal{S}$).}
Measures the proportion of agents converging to the correct answer:
\begin{equation}
\mathcal{S} = \frac{1}{N} \sum_{i=1}^{N} \mathbbm{1}[\hat{y}_i = y^*]
\end{equation}
A task instance is \textit{successful} when $\mathcal{S} = 1$, indicating unanimous convergence.

\paragraph{Partial Correctness Score ($\mathcal{P}$).}
Binary success rate can understate partial progress. 
We introduce a continuous measure of answer quality tailored to each task category: for Level-I, $\mathcal{P}$ is the fraction of agents within tolerance of the ground truth; for Level-II, the fraction of correctly computed elements per local segment; for Level-III, the longest correctly ordered subsequence relative to total length. 
Letting $q_i \in [0,1]$ denote the per-agent quality score:
\begin{equation}
\mathcal{P} = \frac{1}{N} \sum_{i=1}^{N} q_i
\end{equation}
Together with $\mathcal{S}$, this score allows us to isolate where coordination breaks down: the gap $\mathcal{P} - \mathcal{S}$ quantifies performance lost specifically at the reasoning-integration stage rather than at the communication stage.

\paragraph{Token Consumption ($\mathcal{C}$).}
Quantifies computational cost per communication round:
\begin{equation}
\mathcal{C} = \frac{\sum_{i=1}^{N} \sum_{r=1}^{R} t_i^{\text{out}}[r]}{R_{max}}
\end{equation}
where $t_i^{\text{out}}[r]$ is the number of output tokens generated by agent $i$ in round $r$, and $R_{max}$ is the max number of rounds executed.

\paragraph{Communication Density ($\mathcal{D}$).}
Captures inter-agent interaction intensity. 
Here $N(N-1)$ is the directed-edge count when each ordered pair exchanges exactly one message; since agents may send multiple messages to the same recipient across rounds, $\mathcal{D} \in [0, +\infty)$:
\begin{equation}
\mathcal{D} = \frac{\sum_{i=1}^{N} m_i}{N(N-1)}
\end{equation}
Values near 0 suggest sparse, targeted exchanges; $\mathcal{D} = 1$ indicates one message per directed pair on average; values exceeding 1 reflect iterative multi-round exchanges. For the SFS protocol (see Appendix~\ref{app:protocols}), $m_i$ counts the number of times other agents successfully read files written by agent $i$, preserving the same ``information actually transferred'' semantics as direct message-passing.

Together, $\mathcal{S}$ and $\mathcal{P}$ measure \textit{what} agents achieve, $\mathcal{C}$ measures \textit{at what cost}, and $\mathcal{D}$ reveals \textit{how} they coordinate.

\subsection{Execution Pipeline}
\label{sec:pipeline}

Given task $\tau = (f, \mathcal{X}, y^*)$ and configuration $(N, P, \mathcal{M})$, the evaluation proceeds through four phases.

\paragraph{Phase 1: Data Partition.}
$\textsc{Partition}(\mathcal{X}, N) \rightarrow \{\mathcal{X}_1, \ldots, \mathcal{X}_N\}$, where $|\mathcal{X}_i| \approx |\mathcal{X}|/N$ ensures equal partition and no agent holds privileged information.

\paragraph{Phase 2: Agent Initialization.}
Each agent $i$ is initialized with model $\mathcal{M}$ and receives $\textsc{Init}(i) \leftarrow (\text{desc}(f, \mathcal{X}_i), P)$, specifying the core task logic, local data, and protocol constraint. 
The prompt provides task-structural guidance while preserving strategic autonomy (see Appendix~\ref{app:prompt_design}).

\paragraph{Phase 3: Collaborative Execution.}
Agents engage in iterative communication for up to $R_{\max}$ rounds. 
All $N$ agents are activated in parallel within each round: they receive incoming messages from the previous round, independently decide on actions, and execute them simultaneously. 
Messages or files written in round $r$ become visible at the start of round $r+1$. 
Execution terminates when all agents submit answers or the round limit is reached.

\paragraph{Phase 4: Metric Computation.}
The four metrics $(\mathcal{S}, \mathcal{P}, \mathcal{C}, \mathcal{D})$ are computed from submitted answers $\{\hat{y}_i\}_{i=1}^{N}$ and recorded communication logs.
\section{Experiments}
\label{sec:experiments}

We systematically evaluate multi-agent coordination across three orthogonal axes: (i) agent scale, (ii) communication protocol, and (iii) language model, yielding a factorial design that covers qualitatively distinct coordination regimes.

\subsection{Experimental Setup}
\label{sec:setup}

Each evaluation instance in \silobench{} is specified by agent scale $N$, communication protocol $P$, and language model $\mathcal{M}$. 
All models are deployed locally with default temperature and 128K context windows.

\paragraph{Agent Scale ($N$).}
We vary team size across $N \in \{2, 5, 10, 20, 50, 100\}$, chosen to probe qualitatively distinct coordination regimes. 
The minimal team ($N=2$) isolates fundamental pairwise coordination without overhead. 
Small groups ($N \in \{5, 10\}$) allow agents to feasibly track all peers simultaneously. 
Medium scale ($N=20$) begins to make exhaustive peer tracking challenging, pushing agents toward selective communication. 
Large scale ($N \in \{50, 100\}$) makes hierarchical or highly selective coordination effectively necessary---and, as our results confirm, largely beyond the reach of current LLMs.

\paragraph{Communication Protocol ($P$).}

\begin{figure}[t]
    \centering
    \includegraphics[width=1.0\columnwidth]{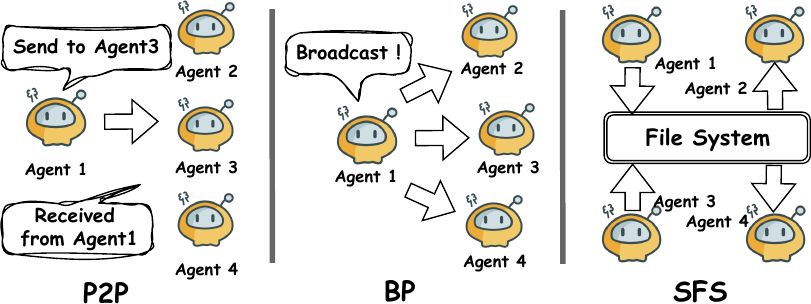}
    \caption{The three communication protocols employed in \silobench{}.}
    \label{fig:protocols}
    
\end{figure}

As shown in Figure~\ref{fig:protocols}, we instantiate three protocols: \textbf{P2P}---directed messaging where agents explicitly address individual recipients; \textbf{BP}---broadcast messaging where each transmission reaches all agents simultaneously; \textbf{SFS}---indirect coordination through a shared file system. Agents retain complete autonomy in deciding what to share, with whom, and when. Detailed specifications are provided in Appendix~\ref{app:protocols}.

\paragraph{Language Model ($\mathcal{M}$).}
All $N$ agents within a configuration share the same model, isolating coordination capability from heterogeneity effects. 
We evaluate three frontier open-source models: \textbf{DeepSeek-V3.1} \citep{liu2024deepseek}, \textbf{GPT-OSS-120B} \citep{openai2025gptoss120bgptoss20bmodel}, and \textbf{Qwen3-Next-80B-A3B} \citep{yang2025qwen3technicalreport}.

\begin{tcolorbox}[colback=lightyellow!10, colframe=black!50, title=Our Experimental Setup]
\texttt{Tasks}: 30 \texttt{(10 per difficulty level)}
\\
\texttt{Scales}: 6 \texttt{(2, 5, 10, 20, 50, 100)}
\\
\texttt{Protocols}: 3 \texttt{(P2P, BP, SFS)}
\\
\texttt{Models}: 3 \texttt{(DeepSeek, GPT, Qwen)}
\end{tcolorbox}

This yields $6 \times 3 \times 3 = 54$ unique configurations and $30 \times 54 = 1{,}620$ total experiments (see Appendix~\ref{app:experiment_details} for infrastructure details). 
To disentangle coordination overhead from intrinsic task difficulty, we additionally conduct \textbf{$N{=}1$ baseline} experiments where a single agent receives the complete global input and answers directly without communication. 
We define \textbf{Relative Coordination Cost (RCC)} $= 1 - \text{SR}(N{=}k)/\text{SR}(N{=}1)$, capturing the fraction of single-agent performance lost to coordination overhead. 
The $N{=}1$ oracle represents the upper bound; \silobench{} asks whether distributed agents can approach this bound through coordination alone.

\subsection{Overall Performance}

Table~\ref{tab:overall_results} summarizes performance across all models and configurations. 
DeepSeek-V3.1 achieves a 36.9\% average success rate, followed by GPT-OSS-120B at 16.9\% and Qwen3 at 8.2\%---a 4.5$\times$ spread. 
Even the strongest model fails nearly two-thirds of the time, establishing that current LLMs cannot reliably coordinate under information silos.

\paragraph{Coordination overhead, not task difficulty, drives the performance gap.}
To confirm that failures stem from coordination rather than intrinsic task hardness, we compare multi-agent success rates against the $N{=}1$ oracle. 
Table~\ref{tab:n1_rcc} reports results for GPT-OSS-120B (trends are consistent across models). 
Even at the smallest team size ($k{=}2$), multi-agent systems already lose 15--49\% of single-agent performance, and RCC compounds steadily with scale, reaching 80--100\% at $k{=}50$ for Level-II and Level-III tasks. 
Crucially, the single-agent success rate difference between Level-I and Level-III is modest---only about 15 percentage points---yet the multi-agent gap balloons to over 18 percentage points, confirming that performance collapse is driven by coordination failure, not by the tasks themselves being harder.

\paragraph{Agents gather information but fail to integrate it.}
While RCC reveals \emph{that} coordination fails, the Partial Correctness Score (PCS) reveals \emph{where}. 
PCS measures continuous answer quality (Section~\ref{sec:metrics}), and the divergence between PCS and SR isolates the reasoning-integration stage as the bottleneck. 
At $N{\geq}50$ on Level-III tasks, SR drops to 0\% while PCS remains at 8--16\%, confirming that agents acquire partial global information but cannot synthesize it correctly.
This dissociation appears even on simpler tasks: averaged across all scales on Level-I tasks, DeepSeek-V3.1 achieves a PCS of 88.0\% yet an SR of only 62.0\% (Table~\ref{tab:overall_results}), a gap of 26 percentage points indicating that agents collectively hold nearly all required information but still fail to produce a correct final answer.

\begin{figure*}[t]
    \centering
    \includegraphics[width=0.95\textwidth]{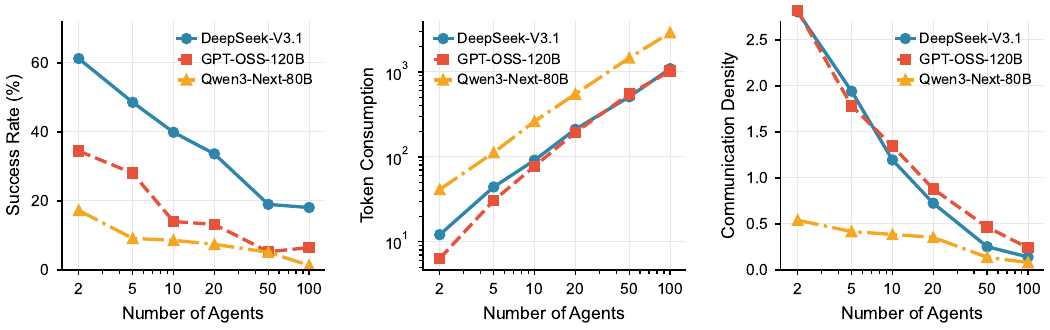}
    \caption{Scaling behavior across agent counts. \textbf{(a)} Success rates decline for all models as team size increases, with sharp drops beyond $N=20$. \textbf{(b)} Token consumption scales roughly linearly with agent count. \textbf{(c)} Communication density decreases at scale, suggesting coordination sparsification.}
    \label{fig:scaling}
\end{figure*}

\begin{figure}[t]
    \centering
    \includegraphics[width=0.95\columnwidth]{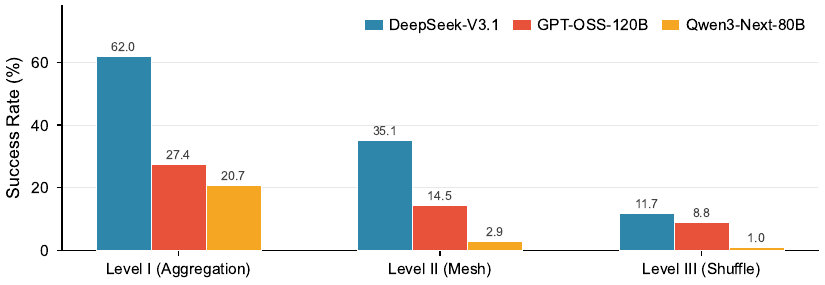}
    \caption{Success rate by difficulty level.}
    \label{fig:difficulty}
\end{figure}

\paragraph{Performance degrades multiplicatively with scale and complexity.}
Figure~\ref{fig:difficulty} and Table~\ref{tab:scale_difficulty} show that task complexity and agent scale interact multiplicatively. DeepSeek-V3.1 drops from 62\% on Level-I to 12\% on Level-III, and Level-III tasks reach \emph{zero} success at $N{\geq}50$, while Level-I tasks remain above 40\% even at 100 agents. 
As Figure~\ref{fig:scaling} illustrates, all models degrade with agent count, and communication density \emph{decreases} at larger scales and agents become sparser in interaction precisely when denser coordination is most needed.

\begin{table*}[t]
\centering
\small
\renewcommand{\arraystretch}{1.1}
\begin{tabular}{l cc cc cc cc cc cc}
\toprule
\multirow{3}{*}{\textbf{Dimension}} 
  & \multicolumn{4}{c}{\textbf{DeepSeek-V3.1}} 
  & \multicolumn{4}{c}{\textbf{GPT-OSS-120B}} 
  & \multicolumn{4}{c}{\textbf{Qwen3-Next-80B-A3B}} \\
\cmidrule(lr){2-5}\cmidrule(lr){6-9}\cmidrule(lr){10-13}
& SR$\uparrow$ & PCS$\uparrow$ & Token$\downarrow$ & Den. 
& SR$\uparrow$ & PCS$\uparrow$ & Token$\downarrow$ & Den. 
& SR$\uparrow$ & PCS$\uparrow$ & Token$\downarrow$ & Den. \\
\midrule
\rowcolor{gray!20} \multicolumn{13}{c}{\textit{By Communication Protocol}} \\
\quad BP  & \textbf{40.4} & \textbf{50.7} & 297.3 & 0.93  
          & 14.4 & 33.1 & 148.7 & 0.78  
          & 4.5  & 12.2 & 846.9 & 0.13 \\
\quad P2P & 38.9 & 50.4 & 363.1 & 1.52  
          & \textbf{20.3} & \textbf{42.6} & 579.8 & 2.25  
          & \textbf{10.5} & \textbf{26.5} & 909.8 & 0.62 \\
\quad SFS & 31.5 & 41.0 & 308.5 & 1.13  
          & 16.0 & 34.7 & 212.8 & 0.74  
          & 9.5  & 18.7 & 864.4 & 0.21 \\
\midrule
\rowcolor{gray!20} \multicolumn{13}{c}{\textit{By Difficulty Level}} \\

\quad Level I   & \textbf{62.0} & \textbf{88.0} & 184.0 & 0.71 
                & \textbf{27.4} & \textbf{56.8} & 187.1 & 1.05 
                & \textbf{20.7} & \textbf{44.4} & 747.1 & 0.62 \\
\quad Level II  & 35.1 & 59.7 & 355.9 & 0.98 
                & 14.5 & 35.3 & 330.1 & 1.18 
                & 2.9  & 11.9 & 990.1 & 0.19 \\
\quad Level III & 11.7 & 27.9 & 439.2 & 1.93 
                & 8.8  & 22.7 & 424.2 & 1.54 
                & 1.0  & \phantom{0}1.5 & 881.6 & 0.16 \\
\midrule
\rowcolor{gray!20} \multicolumn{13}{c}{\textit{By Agent Scale}} \\

\quad N = 2  & \textbf{61.2} & \textbf{78.4} & 12.1  & 2.80 
             & \textbf{34.4} & \textbf{52.0} & 6.3   & 2.82 
             & \textbf{17.2} & \textbf{23.7} & 41.2  & 0.54 \\
\quad N = 5  & 48.5 & 68.2 & 44.2  & 1.94 
             & 28.0 & 47.9 & 30.6  & 1.78 
             & 9.1  & 15.7 & 112.4 & 0.42 \\
\quad N = 10 & 39.9 & 59.1 & 91.3  & 1.19 
             & 14.0 & 36.8 & 77.6  & 1.35 
             & 8.6  & 19.0 & 261.4 & 0.38 \\
\quad N = 20 & 33.6 & 60.2 & 211.0 & 0.72 
             & 13.2 & 37.0 & 194.1 & 0.88 
             & 7.4  & 20.1 & 549.6 & 0.35 \\
\quad N = 50 & 19.0 & 46.8 & 510.3 & 0.25 
             & 5.2  & 27.3 & 549.8 & 0.46 
             & 5.1  & 9.1  & 1466.3& 0.14 \\
\quad N = 100& 18.1 & 46.5 & 1093.8& 0.14 
             & 6.4  & 28.7 & 1024.2& 0.24 
             & 1.3  & 5.1  & 2901.1& 0.08 \\
\midrule
\textbf{Average} & \textbf{36.9} & \textbf{47.1} & 323.0 & 0.82 
                 & 16.9 & 38.3 & 313.8 & 1.01 
                 & 8.2  & 19.8 & 873.6 & 0.25 \\
\bottomrule
\end{tabular}
\caption{Overall performance summary across all models and configurations. SR = Success Rate (\%), PCS = Partial Correctness Score (\%), Token = Token Consumption (tokens/round), Den.\ = Communication Density. Best results per section in \textbf{bold}.}
\label{tab:overall_results}
\end{table*}

\begin{table*}[t]
\centering
\small
\renewcommand{\arraystretch}{1.15}
\setlength{\tabcolsep}{5pt}
\begin{tabular}{ll
  c c >{\columncolor{pink!20}}c  
  c c >{\columncolor{pink!20}}c  
  c c >{\columncolor{pink!20}}c} 
\toprule
& & \multicolumn{3}{c}{\textbf{Level I (Aggregation)}} 
& \multicolumn{3}{c}{\textbf{Level II (Mesh)}}
& \multicolumn{3}{c}{\textbf{Level III (Global Shuffle)}} \\
\cmidrule(lr){3-5}\cmidrule(lr){6-8}\cmidrule(lr){9-11}
\textbf{Scale $k$} & \phantom{x}
& SR($N$=1) & SR($N$=$k$) & RCC
& SR($N$=1) & SR($N$=$k$) & RCC
& SR($N$=1) & SR($N$=$k$) & RCC \\
\midrule
k = 2   && 96.7 & 82.0 & \cellcolor{pink!20}15.2\% & 90.0 & 62.0 & \cellcolor{pink!20}31.1\% & 80.0 & 41.0 & \cellcolor{pink!20}48.8\% \\  
k = 5   && 93.3 & 65.0 & \cellcolor{pink!20}30.3\% & 70.0 & 47.0 & \cellcolor{pink!20}32.9\% & 73.3 & 22.0 & \cellcolor{pink!20}70.0\% \\  
k = 10  && 76.7 & 51.3 & \cellcolor{pink!20}33.1\% & 73.3 & 22.0 & \cellcolor{pink!20}70.0\% & 60.0 &  9.0 & \cellcolor{pink!20}85.0\% \\  
k = 20  && 63.3 & 48.0 & \cellcolor{pink!20}24.2\% & 36.7 & 14.0 & \cellcolor{pink!20}61.8\% & 43.3 &  7.0 & \cellcolor{pink!20}83.8\% \\  
k = 50  && 33.3 & 18.0 & \cellcolor{pink!20}45.9\% & 30.0 &  6.0 & \cellcolor{pink!20}80.0\% & 26.7 &  0.0 & \cellcolor{pink!20}100\% \\    
k = 100 && 20.0 & 10.0 & \cellcolor{pink!20}50.0\% & 13.3 &  5.0 & \cellcolor{pink!20}62.4\% & 10.0 &  0.0 & \cellcolor{pink!20}100\% \\    
\bottomrule
\end{tabular}
\caption{Single-agent baseline SR (\%), multi-agent SR (\%), and Relative Coordination Cost (RCC $= 1 - \text{SR}(N{=}k)/\text{SR}(N{=}1)$) for GPT-OSS-120B across difficulty levels and scales. RCC columns (shaded) quantify the fraction of single-agent performance lost to coordination overhead. Trends are consistent across all three models; full results in Appendix~\ref{app:detailed}.}
\label{tab:n1_rcc}
\end{table*}

\begin{table}[t]
\centering
\small
\renewcommand{\arraystretch}{1.1}
\begin{tabular}{l cccccc}
\toprule
\textbf{Level} & \textbf{N=2} & \textbf{N=5} & \textbf{N=10} & \textbf{N=20} & \textbf{N=50} & \textbf{N=100} \\
\midrule
I & 85.0 & 72.0 & 68.7 & 65.7 & 38.1 & 40.6 \\
II & 61.7 & 55.3 & 28.3 & 29.5 & 17.4 & 14.3 \\
III & 36.2 & 17.2 & 10.0 & 5.7 & 0.0 & 0.0 \\
\midrule
\textbf{Avg} & 61.2 & 48.5 & 39.9 & 33.6 & 19.0 & 18.1 \\
\bottomrule
\end{tabular}
\caption{Success Rate (\%) by agent count and difficulty level for DeepSeek-V3.1 (averaged across all protocols).}
\label{tab:scale_difficulty}
\end{table}

\subsection{Protocol Suitability}

\begin{figure}[ht]
    \centering
    \includegraphics[width=0.95\columnwidth]{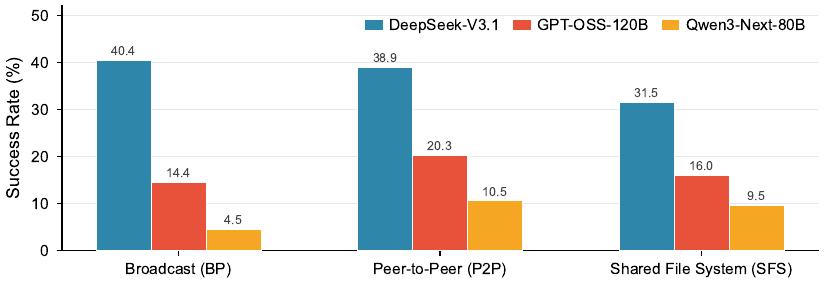}
    \caption{Success rate by communication protocol.}
    \label{fig:protocol}
\end{figure}

Having established that coordination fails broadly, we examine whether protocol choice modulates this failure. 
Figure~\ref{fig:protocol} reveals distinct model-protocol affinities. 
DeepSeek-V3.1 prefers broadcast messaging (40\% with BP vs.\ 32\% with SFS), while GPT-OSS-120B performs best with targeted communication (20\% under P2P vs.\ 14\% under BP), suggesting that protocol suitability depends on how a model balances the cognitive cost of addressing decisions against the noise of undifferentiated broadcasts. 
SFS underperforms in most cases: despite comparable information transfer volume to BP, it consistently yields lower SR, indicating that the bottleneck lies in reasoning about shared state rather than in communication volume.

\section{Analysis and Discussion}
\label{sec:analysis}

The preceding results establish that coordination broadly fails, that failures scale with complexity, and that agents accumulate partial information they cannot synthesize. 
We now investigate the mechanisms: first asking whether agents at least discover the \emph{right structural approach}, then tracing exactly where execution breaks down.

\subsection{Case Study: Emergent Coordination Patterns}

\begin{figure*}[t]
    \centering
    \includegraphics[width=1.0\textwidth]{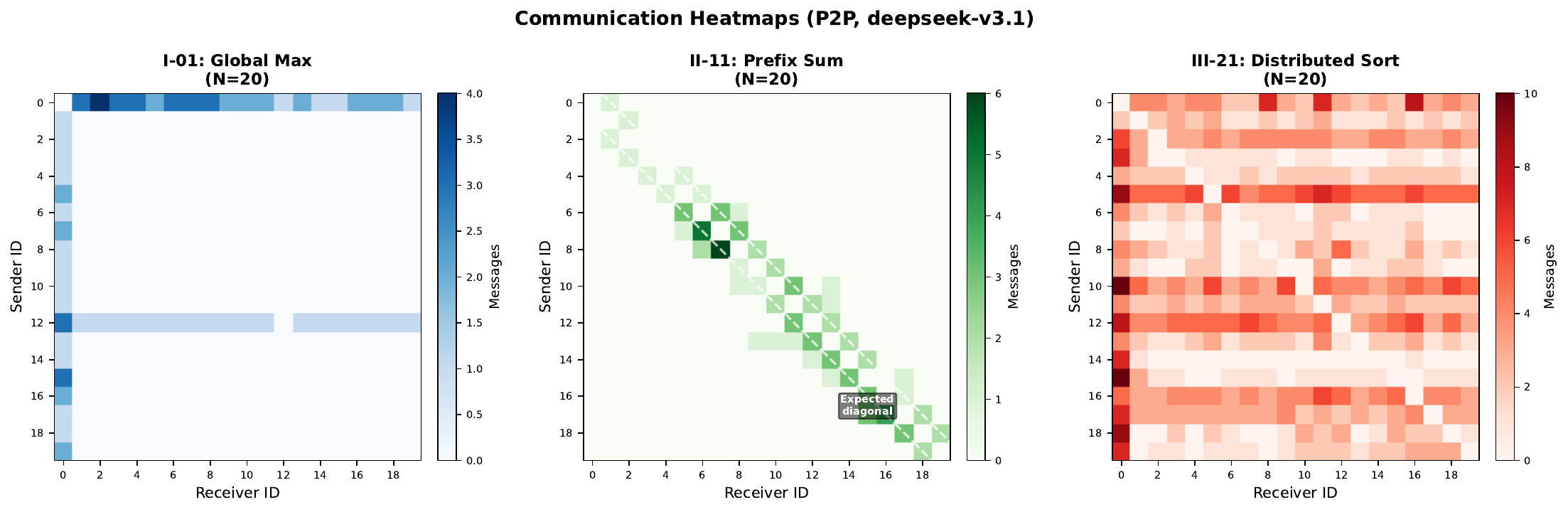} 
    \caption{Communication heatmaps for representative tasks with $N=20$ agents using DeepSeek-V3.1. 
    \textbf{Left}: I-01 (Global Max) shows emergent leader pattern. 
    \textbf{Center}: II-11 (Prefix Sum) exhibits diagonal pattern indicating partial spatial locality discovery. 
    \textbf{Right}: III-21 (Distributed Sort) shows dense all-to-all communication.}
    \label{fig:heatmaps}
\end{figure*}

Figure~\ref{fig:heatmaps} visualizes the communication patterns that agents spontaneously adopt for three representative tasks. 
In the Global Max heatmap (Level-I, left), nearly all message traffic flows into column 0: Agent 0 emerges organically as a central aggregator, producing a near-perfect star topology. 
This self-organized structure closely matches the theoretically optimal pattern and yields high task success. 
In the Prefix Sum heatmap (Level-II, center), a prominent diagonal band reflects agents communicating primarily with their immediate neighbors, correctly capturing the sequential dependency of the prefix computation. 
However, off-diagonal scatter reveals that agents also broadcast beyond their neighbors, generating redundant overhead rather than the clean chain the task requires. 
In the Distributed Sort heatmap (Level-III, right), the matrix is uniformly dense: agents exchange messages with nearly every other agent, which is precisely what global data reorganization demands, yet the high density comes with highly uneven per-agent loads, and some senders dominate entire rows, suggesting uncoordinated flooding rather than structured exchange.

Taken together, these patterns confirm that agents can translate high-level task descriptions into broadly appropriate coordination topologies without explicit instruction. 
Yet the heatmaps also reveal a consistent gap between structural intent and execution quality: even when the right topology emerges, agents over-communicate, distribute load unevenly, or fail to adhere to the optimal pattern. 
This raises the core question addressed next: given that agents communicate in approximately the right way, why do they still so often fail?

\subsection{The Communication-Reasoning Gap}

To classify failure systematically, we apply a two-stage hybrid procedure: rule-based detection identifies \textit{Premature Submission} (agent submits before reaching the task-specific minimum peer count), \textit{Consensus Failure} ($|\{\hat{y}_i\}_{i=1}^N| > 1$), and \textit{Computation Error} (full data receipt confirmed in log but answer incorrect). 
Two independent annotators then reviewed stratified runs, achieving Cohen's $\kappa = 0.87$; disagreements were resolved by discussion.

Analyzing execution logs under this scheme, we identified three distinct failure modes (Table~\ref{tab:failure_modes}). 
\textit{Premature Submission} (37.2\%) is the most prevalent: agents submit before gathering sufficient information---Agent-77 in Task I-06, for instance, submitted after contacting only 28 of 100 agents, yielding answer 208 vs.\ the expected 114. 
\textit{Consensus Failure} (29.9\%) occurs when agents communicate actively but cannot converge; one 100-agent XOR checksum run produced 12 distinct answers, with 86 agents converging on 146 while the remainder submitted values ranging from 42 to 238. 
\textit{Computation Error} (28.6\%) arises when agents collect all required data yet compute incorrectly, such as submitting 619 instead of 620 due to an off-by-one error during final aggregation. 
These modes frequently co-occur: 67 runs exhibit both Premature Submission and Consensus Failure, as early exits prevent subsequent consensus-building and widen the convergence gap for remaining agents.

Together, these three modes define the \textit{Communication-Reasoning Gap}: agents exhibit proficiency in the social mechanics of coordination---formatting messages, responding to peers, organizing information flow---while failing at the computational core of determining information sufficiency and synthesizing distributed state. 
This is not a failure of effort: behavioral comparison shows successful runs complete in fewer rounds, while verification behaviors appear in over 95\% of runs regardless of outcome. The bottleneck is reasoning quality at the integration stage, not communication intent.

\subsection{Implications and Future Directions}

The analyses jointly reveal a structural asymmetry with practical consequences. Coordination overhead does not merely reduce parallelization gains.
For Level-III tasks at $N{\geq}50$, it eliminates them entirely, leaving a coordinated team outperformed by a single agent with full data access. 
Perhaps most counterintuitively, spontaneous leader emergence which is conventionally assumed to help, actively hurts performance on Level-III tasks, because the aggregator becomes overwhelmed by the volume of global data it must process.

Three directions follow. 
First, agents need mechanisms to detect information sufficiency before committing to a final answer. 
Second, the explicit synchronization checkpoints present in successful runs should be formalized as consensus protocols. 
Third, adaptive protocol selection based on task structure could unlock model-protocol co-optimization, given the model-dependent affinities observed. \silobench{} provides the evaluation foundation for tracking progress along all three.

\begin{table}[t]
\centering
\small
\begin{tabular}{lcc}
\toprule
\textbf{Failure Mode} & \textbf{Count} & \textbf{Percent} \\
\midrule
Success & 153 & 50.8\% \\
Premature Submission & 112 & 37.2\% \\
Consensus Failure & 90 & 29.9\% \\
Computation Error & 86 & 28.6\% \\
\bottomrule
\end{tabular}
\caption{Failure mode distribution (categories not mutually exclusive).}
\label{tab:failure_modes}
\vspace{-0.3cm}
\end{table}

\section{Conclusion}
\label{sec:conclusion}

We introduce \silobench{} to evaluate distributed coordination in multi-agent LLM systems across 1,620 experiments. 
The results are unambiguous: current LLMs cannot reliably escape their information silos through coordination alone.

The \textit{Communication-Reasoning Gap} identifies the precise fault line: agents are competent communicators but poor distributed reasoners. 
They spontaneously form task-appropriate topologies and exchange information actively, yet consistently fail to integrate what they have gathered---a dissociation made concrete by the PCS--SR divergence and the RCC analysis showing that coordination overhead eliminates parallelization gains entirely at high complexity. 
Most strikingly, spontaneous leader emergence actively hurts performance on complex tasks, revealing that self-organized centralization creates bottlenecks rather than resolving them.

Closing this gap will require mechanisms for information sufficiency detection, explicit consensus protocols, and adaptive coordination strategies. 
\silobench{} provides the evaluation infrastructure to track progress along these directions.

\section*{Limitations}
While \silobench{} provides a comprehensive framework for evaluating multi-agent collaboration, it has several limitations. Our evaluation only covers three fundamental communication protocols and does not include other coordination mechanisms such as hierarchical protocols, gossip-based dissemination and hybrid approaches. We adopt agent configurations with uniform underlying models, whereas real-world multi-agent systems usually involve heterogeneous compositions with distinct coordination patterns. Closed-source models are not evaluated in this work due to their high cost at our scale and unverifiable, incomparable reported token usage. In addition, our assessment focuses on three frontier LLMs, which may not capture the full spectrum of failure modes across all LLMs since each model has unique characteristics in reasoning logic, communication strategies and error propagation that lead to distinct performance limitations.
\bibliography{custom}
\clearpage
\appendix

\section{Communication Protocols}
\label{app:protocols}

This appendix provides complete specifications for the three communication protocols implemented in \silobench{}. Each protocol defines a distinct coordination substrate constraining the \textit{mechanism} of information exchange while preserving full agent autonomy over \textit{content} and \textit{strategy}.

\subsection{Protocol Overview}

Table~\ref{tab:protocol_comparison} summarizes the three protocols. They span the spectrum of coordination paradigms along three axes: (1) \textit{explicit vs.\ implicit addressing}---P2P requires agents to name recipients, BP eliminates addressing entirely, SFS routes coordination through shared state; (2) \textit{direct vs.\ indirect communication}---P2P and BP involve direct message exchange, SFS agents never ``speak'' to each other explicitly; (3) \textit{default density}---P2P encourages sparse targeted exchanges, BP defaults to dense all-to-all dissemination, SFS density depends entirely on read/write behavior.

\begin{table*}[htbp]
\centering
\small
\renewcommand{\arraystretch}{1.15}
\begin{tabularx}{\textwidth}{@{}l >{\raggedright\arraybackslash}X >{\raggedright\arraybackslash}X@{}}
\toprule
\textbf{Protocol} & \textbf{Description} & \textbf{Available Actions} \\
\midrule
P2P & Directed messaging via agent-addressed mailboxes & \texttt{send\_message}, \texttt{receive\_messages}, \texttt{wait}, \texttt{submit\_result} \\
\addlinespace
BP & Broadcast messaging to all agents simultaneously & \texttt{broadcast\_message}, \texttt{receive\_messages}, \texttt{list\_agents}, \texttt{wait}, \texttt{submit\_result} \\
\addlinespace
SFS & Coordination through shared key-value storage & \texttt{list\_files}, \texttt{read\_file}, \texttt{write\_file}, \texttt{delete\_file}, \texttt{wait}, \texttt{submit\_result} \\
\bottomrule
\end{tabularx}
\caption{Comparison of communication protocols in \silobench{}.}
\label{tab:protocol_comparison}
\end{table*}

\subsection{Peer-to-Peer Protocol (P2P)}
\label{app:p2p}

The P2P protocol implements directed messaging through SQLite-backed mailboxes. Each agent maintains a private inbox; messages are delivered asynchronously to the recipient's buffer until their next activation. Agents must decide not only \textit{what} to communicate but \textit{whom} to contact, enabling evaluation of task-appropriate routing strategy discovery. Available actions are: \texttt{send\_message(target\_id, content)}, which delivers a message to the specified agent; \texttt{receive\_messages()}, which retrieves all pending messages; \texttt{wait()}, which signals completion of the agent's decision for the current round (see below); and \texttt{submit\_result(answer)}, which submits the final answer. Messages are stored in an in-memory SQLite database recording sender ID, recipient ID, content, timestamp, and read status. Delivery ordering within each sender-recipient pair is guaranteed; no global ordering is enforced.

\subsection{Broadcast Protocol (BP)}
\label{app:BP}

The BP protocol implements broadcast messaging where each transmission reaches all other agents simultaneously. An implicit aggregator collects messages each round and distributes the compiled history to all participants. Available actions are: \texttt{broadcast\_message(content)}, \texttt{receive\_messages()}, \texttt{list\_agents()}, \texttt{wait()}, and \texttt{submit\_result(answer)}. Broadcast messages are stored centrally, tagged with sender ID and timestamp, and delivered as a chronologically ordered compiled view at each round.

\subsection{Shared File System Protocol (SFS)}
\label{app:sfs}

The SFS protocol implements indirect coordination through a shared key-value store visible to all agents. Rather than exchanging messages directly, agents read and write to a common namespace, enabling asynchronous coordination analogous to blackboard architectures. Available actions are: \texttt{list\_files(prefix)}, \texttt{read\_file(path)}, \texttt{write\_file(path, content)}, \texttt{delete\_file(path)}, \texttt{wait()}, and \texttt{submit\_result(answer)}. The shared file system is backed by an in-memory SQLite database storing path, content, creation time, and last modification time. Writes are immediately visible to subsequent reads; concurrent writes to the same path follow last-writer-wins semantics.

\subsection{The \texttt{wait()} Action: Formal Specification}
\label{app:wait_spec}

The \texttt{wait()} action is semantically uniform across all three protocols and serves as an explicit round-boundary signal. When an agent invokes \texttt{wait()}, all remaining operations in its current round are skipped, and the agent's decision phase for round $r$ is marked complete. The agent is then suspended until the start of round $r{+}1$.

At the start of round $r{+}1$, the following activation and message-delivery rules apply:

\begin{itemize}[leftmargin=*,nosep]
    \item \textbf{P2P and BP}: The agent's inbox is populated with all messages sent to it (P2P) or broadcast by any agent (BP) during round $r$. These are delivered atomically at round start; no message sent in round $r$ is visible before round $r{+}1$.
    \item \textbf{SFS}: The agent observes the full shared file system state as of the end of round $r$, including all writes committed by other agents during round $r$.
    \item \textbf{No blocking}: \texttt{wait()} does not block on a specific event or agent. It is a ``pass'' that yields control to the synchronous round scheduler. If an agent never calls \texttt{wait()} explicitly, the runtime automatically advances it to the next round after its action budget is exhausted.
    \item \textbf{Post-submit behavior}: Once an agent has called \texttt{submit\_result()}, subsequent rounds are no-ops for that agent---it neither receives new messages nor is activated again.
\end{itemize}

This synchronous round-based model ensures that all $N$ agents observe a consistent snapshot of the communication state at each round boundary, making execution reproducible and analysis tractable.

\section{Prompt Design: Structural Guidance without Role Prescription}
\label{app:prompt_design}

This appendix details the prompt templates used to initialize agents in \silobench{}. Our core design philosophy is to provide \textit{task-structural information} while preserving \textit{strategic autonomy}: prompts convey high-level dependency patterns and potential coordination approaches, but do not prescribe mandatory execution sequences or assign semantic roles.

\subsection{Base Prompt Template}

Each agent receives an initialization prompt following this structure:

\begin{tcolorbox}[colback=lightblue!10, colframe=black!50, title=Agent Initialization Prompt]
\small
\texttt{You are Agent \{agent\_id\} in a multi-agent system consisting of \{N\} agents (IDs range from 0 to \{N-1\}).}

\texttt{\textbf{Task Description:}}\\
\texttt{\{task\_description\}}

\texttt{\textbf{Your Local Data:}}\\
\texttt{\{data\_shard\}}

\texttt{\textbf{Communication Protocol:}}\\
\texttt{\{protocol\_description\}}

\texttt{\textbf{Available Actions:}}\\
\texttt{- \{protocol\_specific\_actions\}}\\
\texttt{- submit\_result(answer): Submit your final answer when confident}

\texttt{Your goal is to coordinate with other agents to compute the globally correct answer. No single agent has sufficient information to solve this task independently. When you have determined the answer, submit it using submit\_result().}
\end{tcolorbox}

The \texttt{\{protocol\_specific\_actions\}} placeholder is instantiated according to the protocol specifications in Appendix~\ref{app:protocols}.

\subsection{Task Description Examples}

Task descriptions convey task structure and potential coordination patterns while leaving concrete implementation decisions to the agents themselves. The three examples below illustrate how descriptions scale from simple aggregation to global reorganization tasks.

\begin{tcolorbox}[colback=lightyellow!10, colframe=black!50, title=Example: Global Maximum (Task I-01)]
\small
\texttt{Find the maximum value across all data distributed among the agents. Each agent holds a portion of a larger array. The correct answer is the single largest integer across the entire distributed dataset. You must coordinate with other agents to determine this global maximum.}
\end{tcolorbox}

\begin{tcolorbox}[colback=lightyellow!10, colframe=black!50, title=Example: Prefix Sum (Task II-11)]
\small
\texttt{Compute the prefix sum array for a sequence distributed across agents. Agent 0 holds elements [0, k), Agent 1 holds elements [k, 2k), and so on. The prefix sum at position i is the sum of all elements from position 0 to i. You must coordinate to compute the correct prefix sum for your portion, accounting for cumulative sums from preceding agents.}
\end{tcolorbox}

\begin{tcolorbox}[colback=lightyellow!10, colframe=black!50, title=Example: Distributed Sort (Task III-21)]
\small
\texttt{Sort the entire distributed array in ascending order. Each agent holds a portion of the unsorted data. The final result should be the complete sorted sequence. You must coordinate to exchange data and determine the correct global ordering.}
\end{tcolorbox}

\subsection{Design Principles}

Our prompts are carefully calibrated to provide structural information without prescribing behavior. On the one hand, descriptions convey whether tasks involve aggregation, sequential dependencies, or global data reorganization (e.g., ``accounting for cumulative sums from preceding agents''); some prompts suggest possible coordination patterns as options rather than requirements (e.g., ``consider establishing a coordinator'' or ``you may exchange with neighbors''); and for complex tasks, prompts may mention general algorithmic paradigms (e.g., ``sample-based partitioning'') without specifying concrete steps. On the other hand, we do not assign semantic roles---no agent is designated ``Manager,'' ``Worker,'' or ``Coordinator''---and prompts never specify ``Step 1: Agent 0 does X, Step 2: Agent 1 does Y.'' Agents decide their own message timing and recipients, and must discover consensus and verification mechanisms independently. The distinction is illustrated below:

\begin{itemize}[leftmargin=*]
    \item[\ding{55}] \textbf{Prescribed (NOT our approach):} ``You are the leader. Step 1: Collect data from all agents. Step 2: Compute the result. Step 3: Broadcast the answer.''
    \item[\ding{51}] \textbf{Our approach:} ``Can you identify a leader to collect and compare results? How would agents coordinate to reach consensus?''
\end{itemize}

All agents receive structurally identical prompts (modulo their ID and data shard), ensuring no agent holds implicit leadership status. The phrase ``No single agent has sufficient information'' is included explicitly to prevent premature submission of partial results. This design tests whether agents can translate high-level task understanding into concrete coordination protocols---a capability that, as our results demonstrate, remains largely absent in current LLMs.

\section{Experimental Details}
\label{app:experiment_details}

\subsection{Agent Scale Rationale}

The six agent counts are chosen to probe qualitatively distinct coordination regimes. The minimal team ($N = 2$) isolates fundamental pairwise coordination without overhead. Small groups ($N \in \{5, 10\}$) allow agents to feasibly track all peers simultaneously. Medium scale ($N = 20$) begins to make exhaustive peer tracking challenging, pushing agents toward selective communication. Large scale ($N \in \{50, 100\}$) makes hierarchical or highly selective coordination effectively necessary---and, as our results confirm, largely beyond the reach of current LLMs.

\subsection{Execution Parameters}

Each configuration is allocated a maximum of $R_{\text{max}} = 100$ communication rounds. Within each round, all agents are activated in parallel: they receive incoming messages from the previous round, independently decide on actions, and execute them simultaneously. Messages or files written in round $r$ become visible to all agents at the start of round $r+1$. An agent exits the coordination loop upon invoking \texttt{submit\_result(answer)}; agents that fail to submit within $R_{\text{max}}$ rounds are assigned a null answer counted as incorrect. Due to computational constraints, each configuration is executed once with fixed random seeds for data generation.

\subsection{Infrastructure}

Experiments were conducted on a GH200 cluster, with up to 50 concurrent configurations executed simultaneously. The compute amounted to approximately 500+ GPU-hours equivalent for each experiment. Complete conversation histories, token counts, and timing information were recorded for all runs.

\subsection{Model Licenses and Intended Use}
\label{app:licenses}

All language models used in this study are open-source and deployed locally on our infrastructure. DeepSeek-V3.1 is released under the MIT License\footnote{\url{https://huggingface.co/deepseek-ai/DeepSeek-V3.1}}, GPT-OSS-120B under the Apache 2.0 License\footnote{\url{https://huggingface.co/openai/gpt-oss-120b}}, and Qwen3-Next-80B-A3B under the Apache 2.0 License\footnote{\url{https://huggingface.co/Qwen/Qwen3-Next-80B-A3B-Instruct}}, all permitting research and commercial use.

\section{Token Budget Feasibility Analysis}
\label{app:token_feasibility}

To verify that \silobench{} operates within practical token limits, we profile token consumption across all 54 configurations, decomposing total usage into three components: the base initialization prompt ($T_{\text{base}}$), the local data shard ($T_{\text{data}}$), and accumulated communication messages ($T_{\text{comm}}$). Table~\ref{tab:token_base_data} summarizes the fixed components, and Table~\ref{tab:token_comm} reports model-dependent communication costs.

\begin{table}[t]
\centering
\small
\renewcommand{\arraystretch}{1.2}
\begin{tabular}{@{}l cccc@{}}
\toprule
\textbf{Component} & \textbf{Min} & \textbf{Mean} & \textbf{Max} & \textbf{Std} \\
\midrule
Base Prompt ($T_{\text{base}}$) & 612 & 748.8 & 989 & 87.3 \\
Data Shard ($T_{\text{data}}$) & 45 & 312.4 & 1,856 & 298.6 \\
\bottomrule
\end{tabular}
\caption{Token consumption of base prompt and data shard per agent (across all configurations).}
\label{tab:token_base_data}
\end{table}

\begin{table}[t]
\centering
\small
\renewcommand{\arraystretch}{1.2}
\begin{tabular}{@{}l cccc@{}}
\toprule
\textbf{Model} & \textbf{Min} & \textbf{Mean} & \textbf{Max} & \textbf{Std} \\
\midrule
DeepSeek-V3.1 & 124 & 8,498.7 & 98,432 & 12,847 \\
GPT-OSS-120B & 89 & 2,049.3 & 45,218 & 5,632 \\
Qwen3-Next-80B-A3B & 156 & 2,299.3 & 52,847 & 6,891 \\
\bottomrule
\end{tabular}
\caption{Communication ($T_{\text{comm}}$) token consumption per agent by model (across all configurations).}
\label{tab:token_comm}
\end{table}

Context window utilization, shown in Table~\ref{tab:context_util}, remains low on average: DeepSeek-V3.1 uses 7.5\% of the 128K budget on average, while GPT-OSS-120B and Qwen3 stay below 3\%. The 95th percentile cases---driven by redundant broadcasting, failed convergence with extended verbose rounds, or agents copy-pasting full message histories---are precisely the coordination inefficiencies \silobench{} is designed to expose. Overall, frontier models (128K--200K context) can run all configurations comfortably; mid-tier models (32K) handle over 90\% of configurations, with Level-III at $N{\geq}50$ potentially requiring truncation; and smaller models (8K) are suitable for $N{\leq}10$ and Level I--II tasks.

\begin{table}[t]
\centering
\small
\renewcommand{\arraystretch}{1.15}
\begin{tabular}{@{}l ccc@{}}
\toprule
\textbf{Model} & \textbf{Mean Util.} & \textbf{95th Pctl.} & \textbf{Max Util.} \\
\midrule
DeepSeek-V3.1 & 7.5\% & 28.4\% & 76.9\% \\
GPT-OSS-120B & 2.4\% & 8.2\% & 35.3\% \\
Qwen3-Next-80B-A3B & 2.6\% & 9.1\% & 41.3\% \\
\bottomrule
\end{tabular}
\caption{Context window utilization (\%) for 128K context models.}
\label{tab:context_util}
\end{table}

\section{Complete Task Specifications}
\label{app:tasks}

Table~\ref{tab:full_mapping} provides the complete mapping between \silobench{} tasks and their algorithmic foundations, including the distributed adaptation approach for each.

\begin{table*}[t]
\centering
\small
\renewcommand{\arraystretch}{1.1}
\begin{tabular}{@{}llp{3.2cm}p{6.5cm}@{}}
\toprule
\textbf{ID} & \textbf{Task Name} & \textbf{Reference} & \textbf{Distributed Adaptation} \\
\midrule
\multicolumn{4}{l}{\textit{Level I: Aggregation (Optimal: Star/Tree Topology, $\mathcal{O}(N)$ messages)}} \\
\midrule
I-01 & Global Maximum & LC-414 & Array partitioned; local max $\to$ global aggregation \\
I-02 & Word Frequency & LC-2085 & Word lists distributed; count target word globally \\
I-03 & Distributed Vote & LC-169 & Vote records partitioned; aggregate to find majority \\
I-04 & Any Match & LC-28 & String collection split; detect pattern in any shard \\
I-05 & Range Count & LC-327 & Count elements in range across shards \\
I-06 & Checksum (XOR) & LC-136 & Data blocks distributed; compute global XOR \\
I-07 & Average Value & LC-1491 & Array partitioned; combine local sums and counts \\
I-08 & Set Union Size & LC-217 & Elements distributed; compute $|\bigcup_i \mathcal{D}_i|$ \\
I-09 & Top-K Selection & LC-215 & Array partitioned; merge local top-K candidates \\
I-10 & Standard Deviation & --- & Two-phase: global mean $\to$ global variance \\
\midrule
\multicolumn{4}{l}{\textit{Level II: Mesh Network (Optimal: Chain Topology, $\mathcal{O}(N)$ messages)}} \\
\midrule
II-11 & Prefix Sum & LC-1480 & Sequential dependency; cumulative offset propagation \\
II-12 & Moving Average & LC-346 & Sliding window spans boundaries; neighbor exchange \\
II-13 & Longest Palindrome & LC-5 & String partitioned; palindromes may cross boundaries \\
II-14 & 1D Life Game & LC-289 & Cellular automaton; boundary cells need neighbor states \\
II-15 & Pattern Search & LC-392 & Subsequence matching across partitioned sequence \\
II-16 & Trapping Rain & LC-42 & Global max-left/max-right propagation required \\
II-17 & Diff Array & LC-1094 & Difference array with boundary handling \\
II-18 & List Ranking & LC-542 & Linked list ranking requires predecessor chain \\
II-19 & Merge Neighbors & --- & Boundary element merging between adjacent agents \\
II-20 & Pipeline Hash & --- & Sequential hash with chained dependencies \\
\midrule
\multicolumn{4}{l}{\textit{Level III: Shuffling (Optimal: Varies, $\mathcal{O}(N \log N)$ to $\mathcal{O}(N^2)$ messages)}} \\
\midrule
III-21 & Distributed Sort & LC-912 & Sample-sort or merge-sort across partitions \\
III-22 & Median of Medians & LC-295 & Iterative median selection across distributed data \\
III-23 & Graph Components & LC-323 & Edges distributed; iterative union-find \\
III-24 & BFS Distance & LC-542 & Graph BFS with distributed edge list \\
III-25 & K-Means Iteration & LC-296 & One K-means iteration with distributed points \\
III-26 & Global Distinct & LC-349 & Hash-based global deduplication \\
III-27 & Collab.\ Filtering & LC-1 & User-item matching with distributed vectors \\
III-28 & PageRank Step & LC-207 & One PageRank iteration with distributed edges \\
III-29 & Load Balance & LC-410 & Task redistribution to minimize load variance \\
III-30 & Matrix Multiply & LC-311 & Row/column partitioned matrix multiplication \\
\bottomrule
\end{tabular}
\caption{Complete specification of \silobench{} tasks with LeetCode references and distributed adaptation descriptions.}
\label{tab:full_mapping}
\end{table*}

\section{Detailed Results}
\label{app:detailed}

\subsection{Task-Level Breakdown for DeepSeek-V3.1}
\label{app:task_breakdown}

Table~\ref{tab:deepseek_detail} provides a comprehensive breakdown of DeepSeek-V3.1's success rate across all 30 tasks and three communication protocols. Tasks achieving $\geq$50\% success rate under any protocol are highlighted in green. Level-I aggregation tasks cluster at the top, with Distributed Vote (I-03) and Any Match (I-04) achieving near-perfect performance across all protocols. Performance degrades sharply for Level-III tasks, with K-Means Iteration (III-25), Collaborative Filtering (III-27), PageRank Step (III-28), and Matrix Multiply (III-30) achieving zero success across all protocols---these tasks require precise numerical computation over all data shards simultaneously, which proves beyond current distributed LLM capabilities.

\begin{table}[t]
\centering
\small
\renewcommand{\arraystretch}{1.05}
\begin{tabular}{l ccc c}
\toprule
\textbf{Task} & \textbf{BP} & \textbf{P2P} & \textbf{SFS} & \textbf{Avg} \\
\midrule
\multicolumn{5}{l}{\textit{Level I: Aggregation}} \\
I-01 Global Max & \cellcolor{gray!30}100 & \cellcolor{gray!30}100 & \cellcolor{gray!30}80 & \cellcolor{gray!30}93.3 \\
I-02 Word Frequency & \cellcolor{gray!30}100 & \cellcolor{gray!30}52 & \cellcolor{gray!30}70 & \cellcolor{gray!30}73.9 \\
I-03 Distributed Vote & \cellcolor{gray!30}100 & \cellcolor{gray!30}100 & \cellcolor{gray!30}100 & \cellcolor{gray!30}100.0 \\
I-04 Any Match & \cellcolor{gray!30}100 & \cellcolor{gray!30}100 & \cellcolor{gray!30}99 & \cellcolor{gray!30}99.6 \\
I-05 Range Count & \cellcolor{gray!30}83 & \cellcolor{gray!30}67 & \cellcolor{gray!30}63 & \cellcolor{gray!30}71.0 \\
I-06 Checksum (XOR) & 17 & 0 & 2 & 6.1 \\
I-07 Average Value & \cellcolor{gray!30}99 & \cellcolor{gray!30}83 & 46 & \cellcolor{gray!30}76.2 \\
I-08 Set Union Size & \cellcolor{gray!30}50 & \cellcolor{gray!30}50 & 42 & 47.5 \\
I-09 Top-K Select & 36 & \cellcolor{gray!30}67 & 21 & 41.2 \\
I-10 Standard Deviation & 17 & 17 & 17 & 16.7 \\
\midrule
\multicolumn{5}{l}{\textit{Level II: Structured Mesh}} \\
II-11 Prefix Sum & \cellcolor{gray!30}84 & \cellcolor{gray!30}80 & 43 & \cellcolor{gray!30}68.8 \\
II-12 Moving Average & 0 & 0 & 0 & 0.0 \\
II-13 Longest Palindrome & 49 & \cellcolor{gray!30}66 & 48 & \cellcolor{gray!30}54.5 \\
II-14 1D Life Game & 27 & 47 & 24 & 32.4 \\
II-15 Pattern Search & 0 & 17 & 17 & 11.1 \\
II-16 Trapping Rain & 33 & 33 & 40 & 35.6 \\
II-17 Diff Array & 48 & \cellcolor{gray!30}62 & 20 & 43.3 \\
II-18 List Ranking & 0 & 0 & 0 & 0.0 \\
II-19 Merge Neighbors & \cellcolor{gray!30}59 & \cellcolor{gray!30}60 & \cellcolor{gray!30}66 & \cellcolor{gray!30}61.8 \\
II-20 Pipeline Hash & \cellcolor{gray!30}52 & 36 & \cellcolor{gray!30}55 & 47.6 \\
\midrule
\multicolumn{5}{l}{\textit{Level III: Global Shuffle}} \\
III-21 Distributed Sort & 33 & 20 & 20 & 24.4 \\
III-22 Median of Medians & 20 & 20 & 0 & 13.3 \\
III-23 Graph Components & 40 & 40 & 14 & 31.3 \\
III-24 BFS Distance & 0 & 0 & 0 & 0.0 \\
III-25 K-Means Iteration & 0 & 0 & 0 & 0.0 \\
III-26 Global Distinct & 33 & 33 & 0 & 22.2 \\
III-27 Collab. Filtering & 0 & 0 & 0 & 0.0 \\
III-28 PageRank Step & 0 & 0 & 0 & 0.0 \\
III-29 Load Balance & 32 & 3 & \cellcolor{gray!30}63 & 32.9 \\
III-30 Matrix Multiply & 0 & 0 & 0 & 0.0 \\
\bottomrule
\end{tabular}
\caption{Success Rate (\%) by task and communication protocol for DeepSeek-V3.1. Tasks with $\geq$50\% success are highlighted in gray background.}
\label{tab:deepseek_detail}
\end{table}

\subsection{Results by Model, Protocol, and Difficulty}

Table~\ref{tab:detailed_all} provides success rates for all model-protocol-difficulty combinations, and Table~\ref{tab:detailed_scale} reports success rates by agent count across models. Together, they confirm that the patterns observed for DeepSeek-V3.1 are consistent across all three models: P2P generally outperforms or matches BP for GPT-OSS-120B and Qwen3, SFS consistently underperforms, and performance degrades monotonically with both complexity level and agent count.

\begin{table}[t]
\centering
\small
\renewcommand{\arraystretch}{1.1}
\begin{tabular}{ll ccc}
\toprule
\textbf{Model} & \textbf{Protocol} & \textbf{L-I} & \textbf{L-II} & \textbf{L-III} \\
\midrule
\multirow{3}{*}{DeepSeek-V3.1}
& BP & 69.7 & 34.5 & 14.7 \\
& P2P & 62.9 & 39.9 & 11.5 \\
& SFS & 53.5 & 30.6 & 9.0 \\
\midrule
\multirow{3}{*}{GPT-OSS-120B}
& BP & 19.6 & 13.9 & 9.7 \\
& P2P & 34.9 & 18.5 & 7.5 \\
& SFS & 27.7 & 11.0 & 9.1 \\
\midrule
\multirow{3}{*}{Qwen3-Next-80B-A3B}
& BP & 9.0 & 3.2 & 1.1 \\
& P2P & 27.6 & 3.4 & 0.3 \\
& SFS & 25.5 & 2.1 & 1.7 \\
\bottomrule
\end{tabular}
\caption{Success Rate (\%) by model, protocol, and difficulty level.}
\label{tab:detailed_all}
\end{table}

\begin{table}[t]
\centering
\footnotesize
\setlength{\tabcolsep}{2pt}
\renewcommand{\arraystretch}{1.0}
\begin{tabularx}{\linewidth}{l *{6}{>{\centering\arraybackslash}X}}
\toprule
\textbf{Model} & \textbf{N=2} & \textbf{N=5} & \textbf{N=10} & \textbf{N=20} & \textbf{N=50} & \textbf{N=100} \\
\midrule
DeepSeek-V3.1 & 61.2 & 48.5 & 39.9 & 33.6 & 19.0 & 18.1 \\
GPT-OSS-120B & 34.4 & 28.0 & 14.0 & 13.2 & 5.2 & 6.4 \\
Qwen3-Next-80B-A3B & 17.2 & 9.1 & 8.6 & 7.4 & 5.1 & 1.3 \\
\bottomrule
\end{tabularx}
\caption{Success Rate (\%) by model and agent count.}
\label{tab:detailed_scale}
\end{table}

\subsection{Communication Density Analysis}

Table~\ref{tab:detailed_density} reports communication density across configurations. A consistent pattern emerges across all models: P2P yields substantially higher densities than BP, reflecting agents' tendency to send multiple targeted messages per pair across rounds; BP densities cluster below 1.0, consistent with one-to-all single broadcasts; and SFS yields notably lower densities than P2P across all models and difficulty levels, indicating that file-based coordination generates sparser cross-agent information flow under our operational definition (read-based transfer counting)---which further explains SFS's systematic under performance on SR despite non-trivial write activity.

\begin{table}[t]
\centering
\small
\renewcommand{\arraystretch}{1.1}
\begin{tabular}{ll ccc}
\toprule
\textbf{Model} & \textbf{Protocol} & \textbf{L-I} & \textbf{L-II} & \textbf{L-III} \\
\midrule
\multirow{3}{*}{DeepSeek-V3.1}
& BP  & 0.56 & 0.82 & 1.46 \\
& P2P & 0.76 & 1.06 & 2.83 \\
& SFS & 0.82 & 1.08 & 1.51 \\
\midrule
\multirow{3}{*}{GPT-OSS-120B}
& BP  & 0.54 & 0.84 & 0.94 \\
& P2P & 1.86 & 2.15 & 2.73 \\
& SFS & 0.73 & 0.54 & 0.95 \\
\midrule
\multirow{3}{*}{Qwen3-Next-80B-A3B}
& BP  & 0.23 & 0.10 & 0.06 \\
& P2P & 1.15 & 0.38 & 0.33 \\
& SFS & 0.48 & 0.07 & 0.09 \\
\bottomrule
\end{tabular}
\caption{Communication Density by model, protocol, and difficulty level.}
\label{tab:detailed_density}
\end{table}

\section{Failure Mode Analysis}
\label{app:failure_analysis}

\subsection{Representative Failure Cases}
\label{app:failure_cases}

Table~\ref{tab:failure_cases} presents representative examples of each failure mode extracted from DeepSeek-V3.1 execution logs. The cases illustrate how failures manifest in practice: premature submission occurs even after reasonable communication volume (Case 4: only 28 of 100 peers contacted before submitting); consensus failure can persist despite near-unanimous agreement, with a single outlier agent preventing full success (Cases 1 and 3); and computation error strikes even when agents have gathered the complete required data (Case 2: off-by-one arithmetic during aggregation).

\begin{table*}[htbp]
\centering
\small
\renewcommand{\arraystretch}{1.2}
\begin{tabularx}{\textwidth}{@{}c >{\raggedright\arraybackslash}p{2.2cm} >{\raggedright\arraybackslash}X >{\raggedright\arraybackslash}X@{}}
\toprule
\textbf{Case} & \textbf{Failure Mode} & \textbf{Description} & \textbf{Key Evidence} \\
\midrule
1 & Consensus Failure & In task I-05, agents communicated extensively but failed to converge, submitting three distinct answers: \{1176, 1182, 1167\}. & 97 of 100 agents submitted 1182; Agent-3 submitted 1176; Agent-55 submitted 1167. \\
\addlinespace
2 & Computation Error & Agent-10 in task I-05 received complete data from all peers but computed an incorrect range count. & Submitted 619 instead of correct answer 620. Arithmetic error during final aggregation. \\
\addlinespace
3 & Consensus Failure & In task I-05 (different instance), 50 agents split between two answers despite communication. & 49 agents submitted 619; Agent-27 submitted 631. \\
\addlinespace
4 & Premature Submission & Agent-77 in task I-06 submitted before collecting sufficient data. & Submitted after receiving data from only 28 of 100 agents. Answer: 208; Expected: 114. \\
\addlinespace
5 & Consensus Failure & 100 agents in task I-06 produced 12 distinct answers for XOR checksum task. & Answers ranged from 42 to 238. Majority (86 agents) converged on 146. \\
\bottomrule
\end{tabularx}
\caption{Representative failure cases from DeepSeek-V3.1 experiments.}
\label{tab:failure_cases}
\end{table*}

\subsection{Failure Mode Definitions and Co-occurrence}

We formally define the three failure modes as follows. \textit{Premature Submission} occurs when an agent invokes \texttt{submit\_result()} before receiving information from a sufficient subset of peers---where ``sufficient'' means the minimum number of agents whose data is required to compute the correct answer. \textit{Consensus Failure} occurs when agents submit multiple distinct answers ($|\{\hat{y}_i\}_{i=1}^{N}| > 1$), indicating that coordination failed to synchronize agents' understanding of global state. \textit{Computation Error} occurs when an agent receives sufficient information but submits an incorrect answer, isolating failures in the reasoning phase from those in the communication phase.

These modes frequently co-occur within single runs, as shown in Table~\ref{tab:failure_cooccurrence}. The high co-occurrence between premature submission and consensus failure (67 cases) suggests a cascading effect: agents submitting early cannot participate in subsequent consensus-building, leaving remaining agents with incomplete information and widening the convergence gap.

\begin{table}[t]
\centering
\small
\renewcommand{\arraystretch}{1.1}
\begin{tabular}{lccc}
\toprule
& \textbf{Premature} & \textbf{Consensus} & \textbf{Compute} \\
\midrule
Premature & 112 & 67 & 45 \\
Consensus & -- & 90 & 52 \\
Compute & -- & -- & 86 \\
\bottomrule
\end{tabular}
\caption{Co-occurrence of failure modes. Diagonal: total occurrences; off-diagonal: joint occurrences.}
\label{tab:failure_cooccurrence}
\end{table}

\subsection{Behavioral Patterns and Leader Emergence}
\label{app:behavior_patterns}

Table~\ref{tab:behavior_patterns_full} compares behavioral metrics between successful and failed runs. Successful runs complete in notably fewer rounds (8.3 vs.\ 12.7), suggesting that effective coordination converges quickly while failed runs engage in extended but ultimately unproductive communication loops. Verification behaviors appear in over 95\% of runs regardless of outcome, confirming that the bottleneck is not communication intent but reasoning quality.

\begin{table}[t]
\centering
\small
\renewcommand{\arraystretch}{1.1}
\begin{tabular}{lcc}
\toprule
\textbf{Metric} & \textbf{Success} & \textbf{Failed} \\
\midrule
Verification Rate & 98.7\% & 95.9\% \\
Strategy Discussion Rate & 93.5\% & 87.2\% \\
Avg.\ Messages per Agent & 31.2 & 27.4 \\
Avg.\ Rounds to Completion & 8.3 & 12.7 \\
\bottomrule
\end{tabular}
\caption{Behavioral comparison between successful and failed runs.}
\label{tab:behavior_patterns_full}
\end{table}

We also examined whether spontaneous leader emergence correlates with task success, classifying an agent as an emergent leader if it receives more than 1.5$\times$ the average number of messages. The results in Table~\ref{tab:leader_emergence_full} are counterintuitive: leader emergence does not consistently improve outcomes, and for Level-III tasks, runs \textit{with} an emergent leader achieve 0\% success versus 33.3\% without one. This suggests that spontaneous centralization at high complexity creates coordination bottlenecks---the designated aggregator becomes overwhelmed by the volume of global data---rather than resolving them.

\begin{table}[t]
\centering
\small
\renewcommand{\arraystretch}{1.1}
\begin{tabular}{lccc}
\toprule
\textbf{Level} & \textbf{Leader Rate} & \textbf{w/ Leader} & \textbf{w/o Leader} \\
\midrule
I & 27.5\% & 56.8\% & 62.1\% \\
II & 21.8\% & 23.5\% & 59.0\% \\
III & 23.8\% & 0.0\% & 33.3\% \\
\bottomrule
\end{tabular}
\caption{Leader emergence rates and associated success rates by complexity level.}
\label{tab:leader_emergence_full}
\end{table}

\subsection{Successful Coordination Examples}
\label{app:success_cases}

To contrast with the failure modes above, we document two illustrative successful patterns. In Case S-1 (Task I-07, $N{=}5$), Agent-0 emerged as coordinator organically: all agents broadcast local results to Agent-0, which computed and rebroadcast the global answer. All agents verified and submitted identically within 4 rounds (100\% success). In Case S-2 (Task I-01, $N{=}10$), agents adopted a distributed verification strategy, with each agent confirming understanding with two neighbors before submission. This redundant verification eliminated consensus failures despite higher message overhead. Both successful patterns share a key property: explicit synchronization checkpoints where agents confirm mutual understanding before proceeding---a discipline entirely absent in failed runs.

\section{Prompt Scaffold Ablation}
\label{app:scaffold_ablation}

To assess the sensitivity of our results to prompt design, we ran a controlled ablation under DeepSeek-V3.1 + P2P with three scaffolding conditions beyond the standard neutral prompt: (a) \textbf{planning round}---a dedicated strategy-discussion round before data exchange begins; (b) \textbf{protocol reminder}---a brief restatement of available communication actions injected at each round; and (c) \textbf{scratchpad hint}---a suggestion to maintain a shared intermediate workspace.

\begin{table}[h]
\centering
\small
\renewcommand{\arraystretch}{1.1}
\begin{tabular}{l ccc}
\toprule
\textbf{Scaffold} & \textbf{L-I SR} & \textbf{L-II SR} & \textbf{L-III SR} \\
\midrule
No scaffold (baseline) & 62.9 & 39.9 & 11.5 \\
+Planning round        & 64.3 & 47.2 & 17.8 \\
+Protocol reminder     & 65.1 & 41.3 & 12.1 \\
+Scratchpad hint       & 63.7 & 44.6 & 14.9 \\
\bottomrule
\end{tabular}
\caption{Prompt scaffold ablation results under DeepSeek-V3.1 + P2P. SR = Success Rate (\%).}
\label{tab:scaffold_ablation}
\end{table}

The planning round yields the most consistent gains ($\sim$5--8\ on Level-II/III); the protocol reminder helps primarily on Level-I; the scratchpad hint benefits intermediate scales ($N{=}10$--20) but cannot prevent collapse at $N{\geq}50$. Critically, qualitative failure patterns remain stable across all conditions: agents continue to communicate actively while failing to translate interaction into correct distributed computation, and the Communication-Reasoning Gap persists regardless of scaffolding. This confirms that the bottleneck reflects genuine LLM limitations in distributed information synthesis rather than a prompting artifact.
\end{document}